
\input harvmac.tex

\voffset=+1.5cm



\font\tenmsy=msbm10
\font\sevenmsy=msbm10 at 7pt
\font\fivemsy=msbm10 at 5pt
\newfam\msyfam 
\textfont\msyfam=\tenmsy
\scriptfont\msyfam=\sevenmsy
\scriptscriptfont\msyfam=\fivemsy
\def\blackB{\fam\msyfam\tenmsy}
\def\Z{{\blackB Z}}

\def\N{{\blackB N}}

\def\frac#1#2{{\textstyle{#1\over #2}}}

\def\b#1{\kern-0.25pt\vbox{\hrule height 0.2pt\hbox{\vrule
width 0.2pt \kern2pt\vbox{\kern2pt \hbox{#1}\kern2pt}\kern2pt\vrule
width 0.2pt}\hrule height 0.2pt}}
\def\ST#1{\matrix{\vbox{#1}}}
\def\STrow#1{\hbox{#1}\kern-1.35pt}
\def\bv{\b{\phantom{1}}}





\def\eqalignD#1{
\vcenter{\openup1\jot\halign{
\hfil$\displaystyle{##}$~&
$\displaystyle{##}$\hfil~&
$\displaystyle{##}$\hfil\cr
#1}}
}

\def\eqalignQ#1{
\vcenter{\openup1\jot\halign{
\hfil$\displaystyle{##}$~&
$\displaystyle{##}$\hfil~&
$\displaystyle{##}$\hfil~&
$\displaystyle{##}$\hfil~&
$\displaystyle{##}$\hfil\cr
#1}}
}

\def\text#1{\quad\hbox{#1}\quad}

\def\la{\lambda}

\def\e{\epsilon}
\def\nuh{{\hat \nu}}
\def\muh{{\hat \mu}}

\def\E{\widehat {E}}
\def\Ac{{\cal A}}

\def\W{\widehat {W}}
\def\rh{{\hat \rho}}
\def\lah{{\hat \lambda}}

\def\y{{\infty}}

\def\rw{\rightarrow}

\def\lra{\leftrightarrow}
\def\su{\widehat{su}}

\def\sp{\widehat{sp}}

\overfullrule=0pt

\newcount\eqnum
\eqnum=0
\def\eq{\eqno(\secsym\the\meqno)\global\advance\meqno by1}
\def\eqlabel#1{{\xdef#1{\secsym\the\meqno}}\eq }

\newwrite\refs
\def\startreferences{
   \immediate\openout\refs=references
   \immediate\write\refs{\baselineskip=14pt \parindent=16pt \parskip=2pt}
}
\startreferences

\refno=0
\def\aref#1{\global\advance\refno by1
   \immediate\write\refs{\noexpand\item{\the\refno.}#1\hfil\par}}
\def\ref#1{\aref{#1}\the\refno}
\def\refname#1{\xdef#1{\the\refno}}

\newcount\exno
\exno=0
\def\Ex{\global\advance\exno by1{\noindent\sl Example \the\exno:

\nobreak\par\nobreak}}

\parskip=6pt



\Title{\vbox{\baselineskip12pt
}}
{\vbox{\centerline{Fusion rules}
	\vskip4pt\centerline{and the}
          \vskip4pt\centerline{Patera-Sharp generating-function
method*}
}}
\footnote{}{*Contribution to the Proceedings of the {\it
Workshop on symmetry in physics in
memory of Robert T. Sharp}, September 12-14, 2002, Centre de
recherches math\'ematiques, Universit\'e
de Montr\'eal, edited by P. Winternitz.  This work was supported in part by
NSERC.}
\vskip.25cm

\centerline{L. B\'egin$^\ddagger$, C. Cummins$^{\sharp}$,  P.
Mathieu$^\natural$ and M.A. Walton$^\dagger$}
\vskip.5cm
\smallskip\centerline{$\ddagger$ \it Secteur Sciences, Campus
d'Edmundston,}
\centerline{\it Universit\'e de Moncton, Nouveau-Brunswick, Canada, E3V 2S8}
\smallskip\centerline{$^\sharp$ \it Mathematics Department,
Concordia University, Montr\'eal, Qu\'ebec, Canada H3G 1M8}
\smallskip\centerline{$^\natural$ \it D\'epartement de Physique,
Universit\'e Laval, Qu\'ebec, Canada G1K 7P4}
\smallskip\centerline{$^\dagger${\it Physics Department,
University of Lethbridge,  Lethbridge, Alberta,
Canada T1K 3M4}}

\noindent

\vskip.75cm
\centerline{{\bf Abstract}}

We review some contributions on fusion rules that were inspired by
the work of Sharp, in particular, the generating-function method for
tensor-product coefficients that he developed with Patera. We also
review the Kac-Walton formula, the concepts of threshold level,
fusion elementary couplings, fusion generating functions and fusion bases.
We try to keep the presentation elementary and exemplify each concept
with the simple $\su(2)_k$ case.


\let\n\noindent
\vfill\break


\newsec{Introduction}

The Patera-Sharp generating function for $su(2)$ tensor products
reads [\ref{J. Patera and R.T. Sharp, in {\it Recent advances in group
theory and
their applications to spectroscopy}, ed. J. Domini (Plenum, New York, 1979);
Lecture Notes in Physics (Springer Verlag, New York, 1979)
vol. 94, p. 175.}\refname\SP]:
$$G^{su(2)}(L,M,N)= {1\over
(1-LM)(1-LN)(1-MN)}\eqlabel\sufun$$
where the multiplicity of the representation $(n)$ in the tensor
product $(\ell)\otimes(m)$ is the coefficient of
$L^\ell M^m N^n$ in the series expansion of (\sufun).
The derivation of this expression is based on manipulations of the character
generating functions
[\ref{R.
Gaskell, A. Peccia and R.T. Sharp, J. Math. Phys. {\bf 19} (1978)
727;  M.
Couture and R.T. Sharp, J. Phys. {\bf A13} (1980) 1925; R.T. Gaskell and R.T.
Sharp, J. Math. Phys. {\bf 22} (1981) 2736;  C. Bodine and R.T.
Gaskell, J. Math.
Phys. {\bf 23} (1982) 2217;   R.V. Moody,
J. Patera and R.T. Sharp, J. Math. Phys. {\bf 24} (1983) 2387;
J. Patera and R.T. Sharp, J. Phys. {\bf A13} (1983) 397;  Y. Giroux,
M. Couture and R.T. Sharp, J. Phys. {\bf
A17} (1984) 715.}\refname\SP].

Key concepts on fusion rules have been obtained by looking for the
affine-fusion extension of this simple-looking
expression and its simplest higher-rank relatives. These are:
the threshold level, fusion
elementary couplings and fusion bases.  Before reviewing these
results, we briefly discuss the
Kac-Walton formula. This last result can also be linked, albeit
loosely,  to Bob Sharp.
Indeed, it is an affine extension of the Racah-Speiser
algorithm for computing tensor-product coefficients, 
one of Sharp's favorite techniques. He
presented it in his course on group theory, where two of the authors
(PM and MW)
learned the fundamentals of this
subject.

\newsec{Fusion rules: the set up}

Fusion rules give the
number of independent couplings between three given primary fields in conformal
field theories. We are interested in those conformal
field theories having a Lie algebra
symmetry.  These are the Wess-Zumino-Witten models [\ref{V.G. Knizhnik and A.B.
Zamolodchikov, Nucl. Phys. {\bf B247} (1984) 83; D. Gepner and E. Witten,
Nucl. Phys. {\bf B278} (1986) 493.}\refname\GW], whose spectrum-generating
algebra is an affine Lie algebra at integer level. Their primary fields are in
1-1 correspondence with the integrable representations of the
appropriate affine
Lie algebra at level
$k$. Denote this set by $P_+^{(k)}$ and a primary field by the corresponding
affine weight $\lah$.  Fusion coefficients ${{\cal 
N}_{\lah\muh}^{(k)}}~^{\nuh}$ are
defined by the fusion product
$$\lah\times \muh = \sum_{\nu\in P_+^{(k)}} {{\cal N}_{\lah\muh}^{(k)}}~^{\nuh}
\; \nuh\eq$$ To simplify the presentation, we consider only the
algebra $\su(N)$.

An affine weight may be written as $$\lah=\sum_{i=0} ^{N-1} \lambda_{i}
{\widehat{\omega}}_{i}=[\lambda_0,\lambda_1,..., \lambda_{N-1}]\eq$$
where
${\widehat{\omega}}_{i}$ denote the fundamental weights of $\su(N)$. If
the Dynkin labels $\la_i$ are nonnegative integers, the weight $\lah$
is the highest
weight of an integrable  representation of $\su(N)$ at level $k$, with
   $k$  defined by $k=\sum_{i=0}^{N-1} \lambda_i$.
To the affine
   weight $\lah$, we associate a finite weight ${\lambda}$ of the finite algebra
$su(N)$:
$${\lambda}=\sum_{i=1}^{N-1} \lambda_i
{\omega}_{i} = (\lambda_1,...,\lambda_{N-1})\eq$$
where ${\omega}_{i}$  are the
fundamental weights of $su(N)$.  
Thus $\lah$ is uniquely fixed by $\la$ and
$k$.

\newsec{Fusion rules and tensor products:  the Kac-Walton formula}

The fusion coefficient ${{\cal N}_{\lah\muh}^{(k)}}~^{\nuh}$ is fixed 
to a large
extent by the tensor-product coefficient pertaining to the product of
the corresponding finite representations.
We denote by ${{\cal N}_{\lambda\mu}}^{\nu}$ the multiplicity of the 
representation
$\nu$ in the tensor product $\la\otimes\mu$:
$$\la\otimes \mu = \sum_{\nu\in P_+} {{\cal N}_{\lambda\mu}}^{\nu}\;  \nu\eq$$
By abuse of notation, we use the same symbol for the highest weight and
the  highest-weight representation.  $P_+$ represents the set of
integrable finite
weights. The precise relation between tensor-product and fusion-rule
coefficients is given by  the Kac-Walton formula [\ref{M.A. Walton, Nucl. Phys.
{\bf B340} (1990) 777; Phys. Lett. {\bf B241} (1990)
365; V.G. Kac, {\it Infinite dimensional Lie algebras}, 3rd
edition (Cambridge Univ. Press, 1990), exercise 13.35.},\ref{ P. Furlan,
A. Ganchev and V.B. Petkova, Nucl. Phys. {\bf B343} (1990) 205;
J. Fuchs and P. van
Driel, Nucl.Phys. {\bf B346} (1990) 632.}]:
$${{\cal N}_{\lah\muh}^{(k)}}~^{\nuh}=
\sum_{{\xi\in P_+
\atop {w\in \W \,,\;\; w\cdot{\hat\xi}=\nuh\in P_+^{(k)}
}}}~{{\cal N}_{\lambda\mu}}^{\xi} {}~\epsilon(w) \eqlabel\kacwal$$ $w$ is
an element of
the affine Weyl group $\W$, of sign $\epsilon(w)$, and the dot
indicates the shifted
action: $w\cdot\lah=w(\lah+\rh)-\rh$, where $\rh$ stands for the affine
Weyl vector
$\rh=\sum_{i=0}^{N-1} {\widehat\omega}_{i}.$

The Kac-Walton formula can be transcribed into a simple algorithm: one first
calculates the tensor product of the corresponding finite weights and then
extends every weight to its affine version at the appropriate level $k$.
Weights with negative zeroth Dynkin label are then
shift-reflected to the integrable affine sector. Weights that
cannot be shift-reflected to the
integrable sector are ignored
(this is the case, for example, for those with zeroth Dynkin label
equal to
$-1$).

Here is a simple example: consider the $su(2)$ tensor-product
$$(2)\otimes
(4)= (2)\oplus(4)\oplus(6)\eq$$ and its affine extension at level 4:
$$[2,2]\times [0,4]= [2,2]+[0,4]+ [-2,6]\eq$$ The last weight must be
reflected since it is not integrable: the shifted action of
$s_0$, the reflection with respect to the  zeroth affine root, is $$s_0\cdot
[-2,6]= s_0 ([-2,6]+[1,1])-[1,1]= [0,4]\eq$$
and this contributes with a minus sign ($\e(s_0)=-1$), cancelling
then the other
$[0,4]$ representation; we thus find: $$[2,2]\times [0,4]= [2,2]\eq$$

The relation between tensor products and fusion was further explored in
[\ref{M.A. Walton, Can. J. Phys. {\bf 72} (1994) 527.}], published in
a special volume of the Canadian Journal of Physics dedicated to
Prof. R.T. Sharp.

\newsec{The idea of threshold level}

The result of the above computation is manifestly level dependent.
Let us reconsider the same product, but at level 5. The affine
extension of the product
becomes
$$[3,2]\times [1,4]= [3,2]+[1,4]+[-1,6]\eq$$
The last weight is thus ignored and the final result is $[3,2]\times [1,4]=
[3,2]+[1,4]$.  For $k>5$ it is clear that there are no truncations, hence no
difference between the fusion coefficients and the tensor products. Moreover,
we see that the representation $(4)$ occurs at level 5 and higher.
    We then say that its {\it threshold
level}, denoted by $k_0$, is
$5$.
The threshold level is thus the smallest
value of $k$ such that the fusion coefficient
${{\cal N}_{\lah\muh}^{(k)}}~^{\nuh}$ is
non-zero, when ${{\cal N}_{\lah\muh}^{(k)}}~^{\nuh}\in \{0,1\}$, for all
levels $k$.\foot{More generally, we say there are
${{\cal N}_{\lambda\mu}}^{\nu}$ couplings, each having its own threshold
level
$k_0$. For fixed $\{\la,\mu;\nu\}$ then, one gets a multi-set of
threshold levels. A simple example:
$su(3)$ with $\la=\mu=\nu$
labelling the adjoint representation has threshold levels $\{2,3\}$.}
If we indicate the threshold level by a subscript, we can write
$$(2)\otimes
(4)= (2)_4\oplus(4)_5\oplus(6)_6\eq$$ To read off a fusion at fixed
level $k$, we
only keep terms with index not greater than $k$.  This implies
directly the inequality
$${{\cal N}_{\lah\muh}^{(k)}}~^{\nuh} \leq
{{\cal N}_{\lah\muh}^{(k+1)}}~^{\nuh} \eqlabel\truc$$ which in turn yields
$$\lim_{k \rightarrow \y} {{\cal N}_{\lah\muh}^{(k)}}~^{\nuh}=
{{\cal N}_{\lambda\mu}}^{\nu}
\eqlabel\limformula$$

The concept of threshold level
was first introduced in  [\ref{C.J.
Cummins, P. Mathieu and M.A. Walton, Phys. Lett. {\bf B254} (1991)
390.}\refname\CMW]. Its origin is directly rooted in the
generating-function method applied to fusion
rules. This is reviewed in the next section, focusing again on the
simple $\su(2)$ case.

\newsec{Fusion generating functions}

The result (\sufun) on the $su(2)$ tensor-product
generating function  can be understood as follows: all couplings can
be described by
appropriate products of three {\it tensor-product elementary couplings}:
$$E_1= LM\qquad E_2= LN\qquad E_3= MN\eq$$
$G^{su(2)}$ can thus be written compactly as
$$G^{su(2)}(L,M,N)= \prod_{i=1}^3{1\over (1-E_i)}\eq$$

How can we construct the affine extension of this generating
function? One certainly needs to
introduce a further dummy variable, say  $d$, in order to keep track
of the extra variable $k$.
Then one could try to introduce factors of $d$ appropriately. A
natural guess is
$$G^{\su(2)}= {1\over
(1-d)(1-dLM)(1-dLN)(1-dMN)}\eqlabel\fgdeux$$
That turns out to be the right answer: this reproduces the $\su(2)_k$
fusion rules.
The prefactor is justified as follows: the fusion of the `vacuum'
with itself, $[k,0]\times
[k,0]=[k,0]$, exists at every level and this is precisely taken into
account by the factor
$1/(1-d)$. With the threshold level insight, we can also naturally
justify the factors of $d$
multiplying the three elementary couplings: their power yields their
threshold level.  This
expression was first proved in [\CMW]. From this example
and that of the $\su(3)$
generating function, we conjectured that any fusion coupling is
characterized by a threshold level.
This was subsequently checked with the $\widehat{so}(5)$ case
[\ref{L. B\'egin, P. Mathieu and M.A.
Walton, J. Phys. A: Math. Gen. {\bf 25} (1992) 135.}\refname\BMW].
This is now understood to be a
consequence of the Gepner-Witten depth rule [\GW], as shown in
[\ref{A.N. Kirillov, P. Mathieu, D.
S\'en\'echal and M. Walton, Nucl. Phys. {\bf B391} (1993) 651.}\refname\KMSW].

  From the above expression, we also infer the existence of {\it fusion
elementary couplings}. For $\su(2)$, there are four of them
$$\eqalignD{ &\E_0 : d~:\quad (0)\otimes(0)\supset(0)_1\qquad
& \E_1 : dLM: (1)\otimes(1)\supset(0)_1,\cr & \E_2 :
dLN: (1)\otimes(0)\supset(1)_1,\qquad & \E_3 : dMN:
(0)\otimes(1)\supset(1)_1.\cr}
\eqlabel\sudeuele$$ As
explained above, subscripts indicate the threshold level.

A re-derivation of (\fgdeux) was presented in
[\ref{L. B\'egin, C. Cummins
and P. Mathieu, J. Math. Phys. {\bf 41} (2000) 7640.}\refname\BCMb].
The method used there
was amenable to generalization, unlike the original proof in
[\CMW]. Consequently, [\BCMb] displays further examples of fusion generating
functions.

\newsec{Tensor products, linear inequalities and elementary couplings}

There are simple combinatorial methods that can be used for calculating
$su(N)$ tensor products, for instance, the Littlewood-Richardson (LR)
rule. It is thus natural to
ask whether  we can read off the threshold level of a coupling from
its LR tableau.

Integrable\ \ weights\ \ in\ \ $su(N)$\ \ can\ \ be\ \ represented\ \
by\ \ tableaux:\ \
the weight
$(\la_1,\la_2,
\ldots ,\la_{N-1})$ is associated to a left justified tableau of $N-1$ rows
with $\la_1+\la_2+\ldots +\la_{N-1}$ boxes in the first row,
$\la_2 + \ldots +\la_{N-1}$
boxes in the second row, etc.  Equivalently, the tableau
has $\la_1$
columns of 1 box, $\la_2$ columns of 2 boxes, etc. The scalar
representation has
no boxes, or equivalently, any number of columns of $N$ boxes. For instance:
$$su(3): (1,1)\;\leftrightarrow\;
\ST{\STrow{\bv\bv}\STrow{\bv}}\qquad su(4):  (2,3,0)
\;\leftrightarrow \;\ST{\STrow{\bv\bv\bv\bv\bv}\STrow{\bv\bv\bv}}\eq$$

The  Littlewood-Richardson
rule is a simple combinatorial algorithm that calculates the decomposition
of the tensor product of two $su(N)$
representations $\la\otimes \mu$.   The second tableau ($\mu$) is filled with
numbers as follows: the first row with
$1$'s, the second row with $2$'s, etc. All the boxes with a $1$ are
then added  to
the first tableau according to
following restrictions: (1) the resulting tableau must be regular:
the number of
boxes in a
given row must be smaller or equal to the number of boxes in the row
just above; (2) the resulting tableau must not contain two boxes
marked by $1$
in the same column.  All the boxes marked by a $2$ are then added to
the resulting tableaux according to the above two rules (with $1$
replaced by $2$) and the further restriction: (3) in counting from right to
left and top to bottom, the  number of
$1$'s must always be greater or equal to the number of $2$'s.
The process is repeated with the boxes marked by a $3, 4, \ldots, N-1$, with
the additional
rule that the number of
$i$'s must always be greater or equal to the number of $i+1$'s when
counted from
right to left and top to bottom.
   The resulting Littlewood-Richardson (LR) tableaux are the Young
tableaux of the irreducible representations occurring in the decomposition.

Here is a simple $su(3)$ example: $(1,1)\otimes(1,1)\supset 2(1,1)$
since we can
draw two LR tableaux with  shape $(1,1)$ and an extra column of three
boxes (the
total number of boxes being preserved, the resulting LR tableau must have 6
boxes):
$$\ST{\STrow{\bv\bv\b1}\STrow{\bv\b2}\STrow{\b1}}
\qquad\ST{\STrow{\bv\bv\b1}\STrow{\bv\b1}\STrow{\b2}}
\eq $$

These rules can be rephrased in an algebraic way as follows [\ref{M. Couture,
C.J. Cummins and R.T. Sharp, J. Phys {\bf A23} (1990)
1929.}\refname\CCS]. Define
$n_{ij}$ to be the number of boxes $i$ that appear in the LR tableau in the row
$j$. The LR conditions read:
$$\lambda_{j-1}+\sum_{i=1}^{k-1} n_{i\,j-1}-\sum_{i=1 }^{k} n_{ij}\geq 0
\quad\quad\quad 1\leq k < j\leq N \quad j\neq 1  \eqlabel\nijrang$$
and
$$\sum_{j=i}^{k} n_{i-1 \, j-1}-\sum_{j=i}^{k} n_{ij} \geq
0 \quad\quad\quad 2\leq i \leq k \leq N \quad {\rm and} \quad i\leq N-1.
\eqlabel\nijlr$$
The weight $\mu$ of the second tableau and the weight $\nu$ of the resulting
LR tableau are easily recovered from these data.

The combined equations (\nijrang) and (\nijlr) constitute a set of linear and
homogeneous inequalities.  We call this the LR (or tensor-product) basis.
As described in [\ref{R.P. Stanley, Duke Math. J. {\bf 40} (1973) 607;
{\it Combinatorics and Commutative Algebra} (Birkhauser,
Boston, 1983).}\refname\Stan],
the Hilbert basis theorem guarantees that every solution can be
expanded in terms of the elementary solutions of these inequalities. This is a
key concept for the following (see [\ref{L.
B\'egin, C. Cummins and P. Mathieu, J.
Math. Phys. {\bf 41} (2000) 7640.}\refname\BCMa] for an extensive
discussion of these methods).
A sum of two solutions translates into the product of the
corresponding couplings,
more precisely, to the {\it stretched product}  (denoted by $\cdot$) of the
corresponding two LR tableaux.  This is defined as follows.  Denote the void
boxes of a LR tableau by a 0, so that
$ n_{0j}
=\sum_{i=j}^{N-1} \lambda_i$,
A tableau is thus  completely characterized by the data $\{n_{ij}\}$ where
now
$i\geq 0$.
Then, the tableau obtained by the stretched product of the
tableaux  $\{n_{ij}\}$ and  $\{ n'_{ij}\}$ is simply described by the numbers
$\{n_{ij}+n'_{ij}\}$, e.g.,
$$\matrix{\ST{\STrow{\bv\bv\b1}\STrow{\b1\b2\b2}\STrow{\b2\b3}\STrow{\b4}
}\cr} \cdot \matrix{\ST{\STrow{\bv\bv\b1}\STrow{\bv\b1\b2}\STrow{\bv\b2}
}\cr}  =
\matrix{\ST{\STrow{\bv\bv\bv\bv\b1\b1}\STrow{\bv\b1\b1\b2\b2\b2}
\STrow{\bv\b2\b2\b3}\STrow{\b4}
   }\cr}\quad  \eq $$

Let us now turn to the $su(2)$ case.
The complete set of inequalities for $su(2)$ variables $\{\la_1, n_{11},
n_{12}\}$ is simply
$$\la_1 \geq n_{12} \qquad
   n_{11}\geq 0 \qquad n_{12}\geq 0\eqlabel\inedeux$$
The first one expresses the fact that two boxes marked by a 1 cannot be in the
same column while the other two are obvious. The other weights are fixed by the
relation $
\mu_1 =n_{11}+n_{12}$ and $
\nu_1=\la_1+n_{11}-n_{12}.$
Any solution of these inequalities describes a coupling.
By inspection, the elementary solutions of this set of inequalities are
$$(\la_1, n_{11}, n_{12}) = (1,0,1), \quad (1,0,0), \quad (0,1,0)\eq$$
(For more complicated cases, we point out that powerful methods to find the
elementary solutions are described in [\BCMb].) These correspond to the
following LR tableaux, denoted respectively
$E_1, E_2, E_3$:
$$E_1: \quad \ST{\STrow{\bv}\STrow{\b1}}\, ,
\qquad E_2: \quad \ST{\STrow{\bv}}\, ,
\qquad E_3: \quad \ST{\STrow{\b1}}\eqlabel\tabdu$$
It is also
manifest
that there are no linear relations between these couplings. Any
stretched product
of these elementary tableaux is an allowed $su(2)$ coupling. Because
there are no
relations between the elementary couplings, this decomposition is
unique. We thus
see that the description of the elementary couplings captures, in a rather
economical way, the whole set of solutions of (\inedeux), that is,
the whole set
of
$su(2)$ couplings.

\newsec{Reformulating the fusion rules in terms of linear inequalitites}

Consider now the affine-fusion extension of the reformulation of the $su(2)$
tensor products in
terms of linear inequalitites. The elementary couplings have a
natural affine extension, denoted by a
hat, and their threshold levels are easily computed from the Kac-Walton
formula.\foot{Let us mention here that for the classical simple Lie algebras,
the tableau methods for tensor products have been modified to implement
the Kac-Walton formula for fusions -- see [\ref{C.J. Cummins and R.C. King,
Can. J. Phys. {\bf 72} (1994) 342.}] in the Sharp volume of the Canadian
Journal of Physics.}
The result is:
$k_0(\E_i)=1$ for
$i=1,2,3$. We observe that these values of  $k_0$ are the same as the number of
columns. Since the product of fusion elementary couplings is also a fusion and
because this decomposition is unique, we can read off the threshold level of
any
coupling, hence of any LR tableau, simply from the number of its columns:
$$ k_0 = \# {\rm columns} = \la_1+n_{11}\eq$$
And since $k$ is necessarily greater that $k_0$, we have obtained the extra
inequality:
$$k\geq \la_1+n_{11}\eqlabel\fudeux$$
This together with  (\inedeux) yield a set of inequalities describing
completely
the fusion rules.  This is what we call a {\it fusion basis}, here the fusion
basis of
$\su(2)$. As in the finite case, the fusion couplings can be described in terms
of elementary fusions. These correspond to the elementary solutions of the four
inequalities, which are easily found to be
$$(k,\la_1, n_{11}, n_{12}) = (1,0,0,0),\quad (1,1,0,1), \quad (1,1,0,0), \quad
(1,0,1,0)\eq$$
They correspond respectively to the coupling
$$\eqalign{ &\E_0 :[1,0]\times [1,0]\supset[1,0]\qquad\quad \E_2: [0,1]\times
[1,0] \supset [0,1],\cr
& \E_1 : [0,1]\times [0,1] \supset [1,0]\qquad\quad  \E_3 : [1,0]\times [0,1]
\supset [0,1].\cr}
\eqlabel\sudeuelex$$ Any fusion has an unique decomposition in terms of these
elementary couplings. For instance
$$[3,2]\times[1,4]\supset[1,4] \quad \lra \quad
\ST{\STrow{\bv\bv\b1\b1\b1}\STrow{\b1}}\quad \lra \quad \E_1\E_2\E_3^3:\quad
k_0=5\eq$$

\newsec{Constructing the fusion basis: Farkas' lemma}

For algebras other than $\su(2)$, the threshold level is not simply
the number of
columns.  So the question is:  how can we derive the fusion basis?
The following strategy was
developed in [\BCMb]:

\item{1)} Write the LR inequalities;

\item{2)} from these, find the tensor-product elementary couplings;

\item{3)} from these, find fusion elementary couplings;

\item{4)} from these, reconstruct the fusion basis.

To go from step 2 to step 3, we need some tools; we describe below a method
based on the outer automorphism group.  Similarly to go from 3 to 4, we need
a further ingredient: this is the Farkas' lemma.  We discuss these
techniques in
turn.

Let us start from the set of tensor-product elementary couplings
$\{E_i, i\in I\}$
for some set $I$ fixed by the particular $su(N)$ algebra under study. For each
$E_i$, we  calculate the threshold level $k_0(E_i)$ and this
datum specifies the affine extension of
$E_i$,  denoted $\E_i$.  We have then a partial set of fusion
elementary couplings
with the set
$\{\E_i,i\in I\}$. Our conjecture is that the missing fusion
elementary couplings
can all be generated by the action of the outer-automorphism group. For
$\su(N)$, this group is simply $\{a^n\,|\, n=0,\ldots, N-1\}$, with
$$a[\la_0,\la_1,\ldots,\la_{N-1}]= [\la_{N-1},\la_0,\ldots,\la_{N-2}]\eq$$
The conjecture is based on the invariance relation
$$ {\cal N}_{a^n\lah, a^m\muh}^{(k)\ a^{n+m}\nuh}=
{{\cal N}_{\lah\muh}^{(k)}}^{~\nuh}\eq$$
It amounts to supposing that the full set is
contained in
$\{ {\cal A }\E_i,  i\in I, \forall \Ac\}$.
Here ${\cal A }\E_i$ indicates a coupling of weights
$${\cal A }\{\lah,
\muh;
\nuh\}= \{a^n\lah,
a^m\muh;
a^{n+m}\nuh\}\, ,\eq$$ $n,m$ being arbitrary integers defined modulo $N$, if
$\E_i$ has weights $\{\lah,
\muh;
\nuh\}$. The
conjectured completeness requires the consideration of all possible pairs
$(n,m)$.\foot{Note that we do not suppose that the action of ${\cal A }$ on an
elementary coupling will necessarily produce another elementary
coupling. Indeed,
the resulting coupling could be a product of elementary couplings.  What is
conjectured here is that all fusion elementary couplings can be
generated in this
way.}

Let us illustrate this with the $\su(2)$ case. Start with the elementary
coupling $E_1:\, (1)\otimes (1)\supset (0)$, which, as already indicated,
arises at level 1: $k_0(E_1)=1$. The corresponding fusion is thus $[0,1]\times
[0,1] \supset [1,0]$,  denoted as $\E_1$.
We now consider  all possible actions of the outer-automorphims group on it.
Since this group is of order 2, there are 4 possible choices for the doublet
$(n,m)$:
$$(a^n,a^m)\in \{(a,a),(1,1), (1,a), (a,1)\}\eq$$
with $a[\la_0, \la_1] = [\la_1, \la_0]$.
This generates the set of four elementary couplings found previously, in the
respective order $\E_0,\E_1,\E_2,\E_3$. Thus, from one tensor-product
elementary
coupling, all four fusion elementary couplings are deduced.

We now turn to Farkas' lemma.  For its presentation, it is convenient to use
an exponential description of the couplings, that is,
$$(k,\la_i,n_{ij})\quad \rw \quad d^kL_i^{k_i}N_{ij}^{n_{ij}}\eq$$
   $d,\,L_i,\,N_{ij}$ being dummy variables.  For instance $\E_1$ is
represented by
$dL_1N_{12}$. If we collectively describe a coupling by the complete set of
variables $\{x_i\}$, we have
$$\{x_i\} \quad \rw \quad \{X_i^{x_i}\}\eq$$
A particular coupling is thus described by a given product $\prod_i
X_i^{x_i}$ with fixed $x_i$.  Its decomposition in terms of elementary
couplings  take the form $\prod_i \E_i^{\,a_i}$.  Now, since $\E_i$ can be
decomposed in terms of the $X_j$ as $$\E_i = \prod_j
X_j^{\e_{ij}}\eqlabel\epde$$ it means that reading off a particular coupling
means that we are interested in a specific choice set of positive integers
$\{x_i\}$ fixed by
$$\sum_i a_i\e_{ij}  = x_j\eq$$
in terms of some positive integers $a_i$.
We are thus looking for the existence
conditions for such a coupling, i.e., the underlying set of linear and
homogeneous inequalities. This is exactly what the Farkas' lemma [\ref{A.
Schrijver, {\it Theory of linear and integer programming} (Wiley,
1986).}] gives us:    given the
knowledge of the elementary couplings, it allows us
to recover the underlying set of
inequalities. For tensor
products, this is of no interest since
we know the corresponding set of inequalities
and our elementary couplings have been extracted from them.  But the
situation is
quite different in the fusion case, where the {\it fusion basis} is unknown.

For our application we need the following modification of the lemma, proved in
[\BCMb]:

\n {\it Lemma}:
Let $A$ be an $r\times m$ integer matrix and let
$\epsilon_j$\  $(j=1,\ldots,n)$ be a set of fundamental
solutions to
$$Ax\geq 0,\quad x\in \N^m. \eqlabel\condB $$ Let
$V$ be the $m\times n$ matrix with entries $V_{i\,j}=(\epsilon_j)_i$\
(for
$i=1,\ldots,m,\  j=1,\ldots,n)$, i.e., the columns
of $V$ are a set of fundamental solutions to (\condB).
Let $e_w$\ $(w=1,\dots,\ell)$ be a fundamental system
of solutions of
$ u^\top V\geq 0$,
(not necessarily positive) $u\in \Z^m$,
and let $E$ be the $\ell\times m$ matrix
with entries $E_{w\,i}=(e_w)_i$, i.e., the rows
of $E$ are the fundamental solutions $e_w$ $(w=1,\dots,\ell)$.
Then the solution set of the system
$$ Ex\geq 0,\quad x\in \N^m \eqlabel\condC$$
is the same as the solution set of (\condB).

To link the lemma to the situation presented above, we note that the entries
$V_{ij}$ of the matrix
$V$ are given here by the numbers $\e_{ji}$ appearing in (\epde).
The relation
$V\, a=x$ describes a generic coupling
$\prod_i\E_i^{\ a_i}$,
and our goal is to find the defining
system of inequalities for $x$ that underly the existence of this coupling.

Take a simple example, the $\su(2)$ case. The
elementary couplings and the corresponding vectors $\e_i$ are
$$\eqalignQ{ &\E_0 : d:\quad & \e_0= (1,0,0,0)\,,\qquad
&\E_2: dL_1:\quad  & \e_2= (1,1,0,0) \cr
&\E_1: dL_1N_{12}:\quad  & \e_1= (1,1,0,1)\,,\qquad
&\E_3: dN_{11} : & \e_3= (1,0,1,0)\cr}\eqlabel\eleuu$$
  From the vectors $\e_{i}$ with components $\e_{ij}$, we form
the matrix $V$ with entries $V_{ij}=\e_{ji}$:
$$V= \pmatrix {1&1&1&1\cr 0&1&1&0\cr 0&0&0&1\cr 0&1&0&0\cr}\eq$$
With $a$ and $x$ denoting the column matrices of entries $a_i$ and $x_i$
respectively, we have the matrix equation
$$V\, a=x\eq$$
Again, this equation describes a general fusion coupling
$\prod_i\E_i^{\ a_i}$.  We now want to unravel the
underlying system of inequalities.  For this, we consider the
fundamental solutions
of
$$u^\top\, V\geq 0\eq$$ where $u$ is the vector of entries $u_i$.
These inequalities read
$$ u_0\geq 0\, ,\qquad
   u_0+u_1+u_3\geq 0\, ,\qquad
u_0+u_1\geq 0\, ,\qquad
u_0+u_2\geq 0\eq$$
In this simple case, the elementary couplings can be found by
inspection and these
are:
$$ e_0=(1,-1,-1,0),\qquad e_1=(0,0,0,1),\qquad
e_2=(0,1,0,-1),\qquad
   e_3=(0,0,1,0)\eq$$
    Finally, we consider the conditions
$e_i\, x\geq 0$, with
$(x_0,x_1,x_2,x_3)= (k,
\la_1, n_{11}, n_{12})$.  They read, in order,
$$ k\geq \la_1+n_{11},\qquad\quad
n_{12}\geq 0,\qquad\quad
\la_1\geq n_{12},\qquad\quad
n_{11}\geq 0\eqlabel\inedeux$$
The last three conditions define the LR basis. The first one is the additional
fusion constraint.  Together, they form the $\su(2)$ fusion basis.

\newsec{Constructing the fusion basis: polytope techniques}

In the previous section, the Farkas' lemma has been used to construct
the fusion basis out of the
set of fusion elementary couplings. There are alternative approaches,
however.  Another one is based
on the reinterpretation of the fusion-rule
computations in terms of counting
points inside a polytope.
    A
polytope can be described by its vertices or its
facets.  In our context, the
{\it vertices} are represented by the fusion elementary
couplings and the {\it facets} are
the inequalities for which the elementary couplings
are the elementary
solutions.
The
reconstruction of the facets of a polytope from  its
vertices is thus another way to generate the fusion basis.  This
method is described in
[\ref{L. B\'egin, C. Cummins, L. Lapointe and P. Mathieu,
J. Math. Phys. {\bf 43} (2002)
4180.}].

In the special case of $su(N)$, Berenstein-Zelevinsky triangles
can also be used to derive the polytope description of a
fusion basis
[\ref{J. Rasmussen and M.A. Walton, Nucl. Phys. {\bf B620} (2002) 537,
J. Phys. {\bf A35} 6939.}], by considering so-called virtual triangles.
Multiple sum formulas can then be written for fusion coefficients of various
types.
However, as are all methods to date, this one is difficult to extend
to higher rank. Assigning a threshold level to a
Berenstein-Zelevinsky triangle becomes very rapidly more difficult with
increasing rank (see [\ref{L. B\'egin,
A.N. Kirillov, P. Mathieu and M. Walton, Lett. Math. Phys. {\bf 28} (1993)
257; L. B\'egin, P. Mathieu and M.A.
Walton, Mod. Phys.
Lett. {\bf A7} (1992) 3255.}\refname\LB]).

\newsec{Conclusion}

The fusion bases have been constructed for
$\su(3)$, $\su(4)$ and $\sp(4)$. Note that for algebras other that
$\su(N)$, we replace the LR basis by the Berenstein and Zelevinsky basis
[\ref{
A.D. Berenstein and A.V.  Zelevinsky, J. Geom. Phys. {\bf 5} (1989)
453.}\refname\BZin].  This leads to explicit expressions for the threshold
levels, hence for the fusion coefficients.\foot{Related  methods for
obtaining the threshold
level are presented in [\LB].}

We stress that the reformulation of the problem of computing fusion rules
in terms of a fusion basis solves, in principle, the quest for
a combinatorial method, since it reduces a fusion computation to solving
inequalities. But it is probably not the optimal solution to the quest
for an efficient combinatorial description.

The main open problem concerning fusion bases is to find
a fundamental and Lie algebraic way of deriving the basis,
analogous in spirit to the
Berenstein-Zelevinsky conjectures for generic Lie algebras in [\BZin].

The methods described in section 8, involving Farkas' lemma, are
general and powerful, but they may not be Lie algebraic enough.
Perhaps one should step back and look at
a first principles description of the tensor product couplings,
and its adaptation to fusion. This was done in
[\ref{J. Rasmussen and M.A. Walton, Nucl. Phys. {\bf B616} (2001) 517.}].
Three-point functions were calculated that can be regarded as
generating functions for tensor product couplings, and
a very simple method
was found for adapting the results to fusion couplings. In principle,
the procedure
works for any semi-simple Lie algebra. Unfortunately,
these more Lie algebraic methods are inevitably more involved. It gives more
information (such as operator product coefficients instead of just fusion
coefficients), but it is not clear that it can be implemented effectively
on higher rank algebras.

\vfill\eject
\centerline{\bf REFERENCES}
\vskip 1cm
\immediate\closeout\refs \vskip 0.5cm
    \message{References}\input references
\vfill\eject

\end

was a favorite one of Sharp and
learned the fundamental of this
discovery of this formula.}

\centerline{Fusion rules}

\smallskip\centerline{and the}

\smallskip\centerline{Patera-Sharp generating-function
method}

\centerline{L. B\'egin$^*$, C. Cummins$^{\sharp}$,  P.
Mathieu$^\natural$ and M.A. Walton$^\dagger$}
\smallskip\centerline{* \it Secteur Sciences, Campus
d'Edmundston, Universit\'e de
Moncton, N.-B., Canada, E3V 2S8}
\smallskip\centerline{$^\natural$ \it D\'epartement de Physique,
Universit\'e Laval, Qu\'ebec, Canada G1K 7P4}
\smallskip\centerline{$^\sharp$ \it Mathematics Department,
Concordia University, Montr\'eal, Qu\'ebec, Canada H3G 1M8}
\centerline{$^\dagger${\it Physics Department,
University of Lethbridge, Lethbridge, Alberta,
Canada\ \ T1K 3M4}}

In a sense, this course has thus been influential in the
discovery of this formula.
to the above